\documentclass[aps,pra,numerical,reprint,showpacs]{revtex4-1}  %aip,jmp     aps,prb
\usepackage[dvips]{graphicx}%
\usepackage{bm}% 
\usepackage{amsmath}
\usepackage{dcolumn}
\usepackage[normalem]{ulem} %solo sirve para tachar texto
\usepackage[colorlinks=true,linkcolor=blue,citecolor=blue ,urlcolor=blue]{hyperref}

\begin{document}

%****** new commands
\newcommand{\op}[1]{{\bm{#1}}}
\newcommand{\bra}{\langle}
\newcommand{\ket}{\rangle}
\newcommand{\new}[1]{\textcolor[rgb]{0,0,0.7}{\uline{#1}}}
\newcommand{\old}[1]{\textcolor[rgb]{1,0,0}{\sout{#1}}}
\newcommand{\nota}[1]{\textcolor[rgb]{1,0,0.5}{{#1}}}
\newcommand{\otranota}[1]{\textcolor[rgb]{0,0.7,0.5}{{#1}}}
%************************

%\preprint{APS/123-QED}

\title{Optimal basis for the generalized Dicke model}
\author{S. Cordero} 
\email{sergio.cordero@nucleares.unam.mx}
\author{E. Nahmad-Achar}
%\email{nahmad@nucleares.unam.mx}
%
\author{O. Casta\~nos}
%\email{ocasta@nucleares.unam.mx}
%
\author{R. L\'opez-Pe\~na}
%\email{lopez@nucleares.unam.mx}
%

%
\affiliation{%
Instituto de Ciencias Nucleares, Universidad Nacional Aut\'onoma de M\'exico, Apartado Postal 70-543, 04510 M\'exico Cd. Mx., Mexico }

\date{\today}

\begin{abstract}
	A methodology is devised for building optimal bases for the generalized Dicke model based on the symmetry adapted variational solution to the problem. At order zero, the matter sector is constructed by distributing $N_a$ particles in all the possible two-level subsystems connected with electromagnetic radiation; the next order is obtained when the states of $N_a-1$ particles are added and distributed again into the two-level subsystems; and so on. In the electromagnetic sector, the order zero for each mode is the direct sum of the Fock spaces, truncated to a value of the corresponding constants of motion of each two-level subsystem; by including contributions of the other modes, the next orders are obtained.  As an example of the procedure we consider $4$ atoms in the $\Xi$ configuration interacting dipolarly with two modes of electromagnetic radiation. The results may be applied to situations in quantum optics, quantum information, and quantum computing.
\end{abstract}

%
%\pacs{}%
%

\maketitle

\section{Introduction}

The study of a finite number  of $n$-level matter systems, be they atoms, artificial atoms, spin systems, or molecules, interacting with an electromagnetic field of $\ell$-modes, has regained importance as applications in quantum optics, quantum gates, and quantum information have become realistic. In particular electromagnetic induced transparency, population trapping, and quantum memories requiere the presence of at least 3 atomic (matter) levels~\cite{haroche06, benenti04, joshi12}.

Dynamically-driven quantum coherence in qubit systems, which are made to cross the quantum phase transition into the superradiant region, has been shown~\cite{benedict96,gomez-ruiz18, gomez-ruiz16}, as well as the generation of field-matter entanglement in the system by varying the light-matter coupling parameter~\cite{acevedo15}.

Even if one restricts the number of photons in the radiation field with some upper bound, a strong limitation in these studies is that the dimension of the Hilbert space becomes unwieldy as the number of atoms and total excitations grow. 

In this work, we build a sequence of ever approximating bases for the infinite-dimensional Hilbert space $\mathcal{H}$ of matter interacting with radiation, in order to carry out a complete study for non-interacting particles exchanging energy with $\ell$ modes of electromagnetic radiation. In particular, an upper bound is placed on the total number of excitations of the system, essentially limiting the number of photons, in order to obtain a finite-dimensional Hilbert space to work on. This upper bound is chosen in such a way that the ground state (which is the one to be studied) obtained in this truncated space differs from the exact ground state by no more than a certain allowed error ${\rm e}_{\rm rr}$ as measured by the fidelity $F$ between the two states. We here show examples for both ${\rm e}_{\rm rr} = 10^{-10}$ and ${\rm e}_{\rm rr} = 10^{-15}$. The value for ${\rm e}_{\rm rr}$ is of course arbitrary, and will be demanded by the type of application to be given. For ${\rm e}_{\rm rr} = 10^{-10}$, for example, the error in the energy surface and observables is less than $10^{-8}$.

The fact that we have an iterative method for reducing a system of $n$-level atoms interacting with radiation to a system of $n-1$ levels~\cite{cordero15}, plus the result that the polychromatic phase diagram divides itself into monochromatic subregions~\cite{cordero15,cordero16,cordero17} where only one of the radiation modes strongly dominates, suggest a methodology for reducing the space dimension even further. This methodology is used to build a sequence of bases for the Hilbert space which approximates better the exact results as we move along the sequence. Previously untractable problems may be tackled in this way, and depending on the desired approximation the appropriate basis may be chosen.

The iterative method  just mentioned allows the study of  the ground state of a very general atomic system of $n$-levels, in the presence of an electromagnetic field of $\ell$-modes, even in the case where each mode produces transitions between more than one pair of levels, to be carried out by studying subsystems consisting of 2 atomic levels interacting dipolarly with one radiation mode. 

The investigation of the 4-level $N$ and $\lambda$ atomic configurations interacting with 2 radiation modes has been shown to present qualitatively different quantum phase diagrams~\cite{cordero16}. While the collective superradiant regime in the latter divides itself into two subregions, corresponding to each of the modes, that of the former may be divided into two or three subregions depending on how the field modes divide the atomic system into 2-level subsystems. Furthermore, a  four-level Josephson circuit shows the dynamics of two-qubit systems~\cite{svetitsky14}. This shows the importance of studying 2-level atomic systems under the influence of one-mode radiation fields (for a review, cf.~\cite{garraway11}).  Recently, the importance of adding unitary invariant phase factors in the matter-field interactions of two- and three-level particles has been established, which can be seen as a canonical transformation represented as a unitary transformation~\cite{dirac82}. They found that the phase factors affect the intrinsic symmetry of the two- and three-level Dicke models interacting with one mode of radiation~\cite{fan16}.  However we want to stress that the ground-state phase diagram presented is very similar to the case of the Dicke model but with two modes of electromagnetic radiation~\cite{cordero15}.

Our methodology shows how to study the properties of the ground state by reducing the full system to 2-level subsystems, and has been tested in the particular case of one 3-level atom~\cite{cordero19} and in the existence of universal parametric curves~\cite{castanos18b}. Here we show that the method is generalizable to any finite number of atoms.

After presenting the general methodology, the study of four atoms in the $\Xi$-configuration in the presence of two electromagnetic modes is given in full as an example. The energy surface and the photon number fluctuations are also calculated.

This paper is organized as follows: In Sec.~\ref{s.general} the generalized Dicke model is presented. Sec.~\ref{s.Fbasisnl} builds the full basis of the Hilbert space, as well as a criterion to obtain convergence of the solution based on the fidelity of states. We show that this procedure yields the minimum number of excitations needed to obtain convergence in a related $2$-level system, and we discuss a method to obtain the convergence in the general $3$-level case. We also calculate the minimum energy surface for the exact quantum solution of a $3$-level system. In Sec.~\ref{s.redB}, a reduction method is presented which results in a sequence of ever-approximating bases for the Hilbert space. Sec.~\ref{s.Xiconf} presents the results for a $3$-level system interacting with two modes of electromagnetic field, obtained from the exact solution and from the reduced bases. These results are discussed and compared. Finally, in Sec.~\ref{conclusion}, some concluding remarks are given. 

\section{Generalized Dicke model}\label{s.general}

Let us consider $N_a$ atoms of $n$-levels interacting dipolarly with $\ell$-modes of electromagnetic field, where the transition between any given pair of atomic levels is promoted only by one mode of the field. The Hamiltonian  is composed of two terms:  a diagonal  part $\op{H}_D$ containing the field and matter sectors, and a non-diagonal $\op{H}_{int}$ containing the matter and field dipolar interactions. So we can write ($\hbar=1$)~\cite{cordero17} 
\begin{eqnarray}\label{eq.Hnl}
\op{H} = \op{H}_D + \sum_{s=1}^{\ell}\op{H}_{int}^{(s)}\,,
\end{eqnarray}
with 
\begin{equation}\label{eq.HD}
\op{H}_D = \sum_{s=1}^\ell \Omega_{s}\,\op{\nu}_{s} + \sum_{k=1}^n\omega_k\,\op{A}_{kk}\,, 
\end{equation}
where $\Omega_s$ denotes the $s$-mode field frequency, $\omega_k$ the frequency of the atomic level $k$, $\op{\nu_s}$ the bosonic field operator  $\op{\nu}_s=\op{a}_s^\dag\,\op{a}_s$ of mode $s$, with $ \op{a}_s^\dag$ and $\op{a}_s$ the creation and annihilation operators, and $\op{A}_{kk}$ and $\op{A}_{jk}$ are atomic weight and transition operators, respectively, obeying the unitary algebra ${\rm U}(n)$ in $n$ dimensions
\begin{equation}
\left[\op{A}_{jk},\op{A}_{lm}\right] = \delta_{kl}\,\op{A}_{jm} - \delta_{jm}\,\op{A}_{lk}\,.
\end{equation}
For the totally symmetric irreducible representation of ${\rm U}(n)$, the generators have a bosonic representation as $\op{A}_{jk}=\op{b}^\dag_j\,\op{b}_k$ and first order Casimir operator
\begin{equation}\label{eq.Nanl}
\sum_{k=1}^n \op{A}_{kk} = N_a\,\op{1}_{\rm matt}\,,
\end{equation}
with $\op{1}_{\rm matt}$ the identity operator in the matter sector of the Hilbert space.

The second contribution term in~(\ref{eq.Hnl}) reads
\begin{equation}\label{eq.Hint}
\op{H}_{int}^{(s)} = -\frac{1}{\sqrt{N_a}}\sum_{j<k}^n \mu_{jk}^{(s)} \left(\op{A}_{jk}+\op{A}_{kj} \right)\left(\op{a}^\dag_{s} + \op{a}_{s}\right)\,,
\end{equation}
where $\mu_{jk}^{(s)}$ is the matter-field coupling parameter and denotes the dipolar intensity. Since we have assumed that  transitions between a pair of atomic levels are promoted only by one mode of electromagnetic field, say $\Omega_s$, one has as condition that if $\mu_{jk}^{(s)}\neq 0$ then $\mu_{jk}^{(s')}=0$ for all $s'\neq s$.

The adopted convention $\omega_j\leq \omega_k$ for $j<k$ on the atomic levels allows us to refer to a particular atomic configuration by the appropriate choice of vanishing dipolar strengths $\mu_{jk}^{(s)}$.  Also, fixing values $\omega_1=0$ and $\omega_n=1$, one may to refer to all energy (and frequency) quantities in units of $\hbar\,\omega_n$ (and $\omega_n$).  

For each mode $s$ the interaction term~(\ref{eq.Hint}) has the form $\op{H}_{int}^{(s)}=\op{R}_{int}^{(s)}+\op{C}_{int}^{(s)}$, with $\op{R}_{int}^{(s)}$ the rotating and $\op{C}_{int}^{(s)}$ the counter-rotating terms. The rotating term preserves the total number of excitations
\begin{equation}\label{eq.HintRWA}
\op{R}_{int}^{(s)} = -\frac{1}{\sqrt{N_a}}\sum_{j<k}^n \mu_{jk}^{(s)} \left(\op{A}_{jk}\,\op{a}_{s}^\dag+\op{A}_{kj} \,\op{a}_{s}  \right)\, ,
\end{equation}
because a decrease (or increase) in an atomic excitation involves an increase (or decrease) in the photon number. The counter-rotating term does not preserve the total number of excitations, and is given by 
\begin{equation}\label{eq.HintNRWA}
\op{C}_{int}^{(s)} = -\frac{1}{\sqrt{N_a}}\sum_{j<k}^n \mu_{jk}^{(s)} \left(\op{A}_{jk}\,\op{a}_{s}+\op{A}_{kj} \,\op{a}^\dag_{s}  \right)\,.
\end{equation}
The Hamiltonian in the rotating wave approximation (RWA), obtained when the counter-rotating term is neglected, would be called the {\it generalized Tavis-Cummings model} (GTCM).

\section{Full Basis}\label{s.Fbasisnl}

A complete basis for the Hamiltonian~(\ref{eq.Hnl}) is formed by the direct product of the Hilbert spaces of the field and matter sectors. An element is of the form
\begin{equation}\label{eq.statenl}
|\vec{\nu};\vec{n}\ket:=|\nu_{1},\,\dots,\,\nu_{\ell};\,a_1,\,\dots,\,a_n\ket\,,
\end{equation} 
which satisfies the eigenvalue equations
\begin{equation}
\op{\nu}_{s}|\vec{\nu};\vec{n}\ket=\nu_{s}|\vec{\nu};\vec{n}\ket\,, \quad  \op{A}_{kk}|\vec{\nu};\vec{n}\ket=a_k|\vec{\nu};\vec{n}\ket\,,
\end{equation}
for the number of photon operator $\op{\nu}_{s}$ of the mode $\Omega_s$ and the particle number operator $\op{A}_{kk}$ for the atomic level $k$. 

Denoting the Fock space of each mode $\Omega_s$ by
\begin{equation}
{\cal F}_{s} := \left\{ |\nu_{s}\ket \bigg| \nu_{s} =0,\,1,\,2,\,\dots\right\}\,,
\end{equation}
with infinite dimension, and the matter space by
\begin{equation}\label{eq.Mnl}
{\cal M} := \left\{ |a_1,\,\dots,\,a_n\ket \bigg| \sum^n_{k=0}a_k=N_a \, , \, a_k \geq 0 \right\}\,,
\end{equation}
with finite dimension given by $\binom{N_a+n-1}{n-1}$ because the number of particles is preserved~(\ref{eq.Nanl}),
the full basis is then
\begin{equation}
{\cal B} := \otimes_{s=1}^\ell{\cal F}_{s}\otimes{\cal M}\,.
\end{equation}

\subsection{Parity Adapted Basis}

\begin{table}
\caption{Coefficients corresponding to the $\op{K}_\zeta$ operators in Eq.~(\ref{eq.Kjnl}), for the $\Lambda$, $\Xi$ and $V$ atomic configurations. Subscripts $s$, $s'$ and $s''$ correspond to the modes of the transitions $1\rightleftharpoons 2$, $1\rightleftharpoons 3$, and $2\rightleftharpoons 3$, respectively.}\label{t.K3l}
\vspace{5mm}
\begin{center}
\begin{tabular*}{\linewidth}{c c | c c c c c c}
Conf. &  $\op{K}_\zeta$ & \phantom{x} $\eta_{s}^{(\zeta)}$ \phantom{x}  & \phantom{x}  $\eta_{s'}^{(\zeta)}$ \phantom{x}  & \phantom{x}  $\eta_{s''}^{(\zeta)}$ \phantom{x}  & \phantom{x}  $\lambda_{1}^{(\zeta)}$ \phantom{x}  &  \phantom{x} $\lambda_{2}^{(\zeta)}$ \phantom{x}  & \phantom{x}  $\lambda_{3}^{(\zeta)}$\phantom{x}  \\[1mm] \hline\hline & & & & & & & \\[-3mm]
$\Lambda$ & & & & & & & \\
&$\op{K}_1$ &  0 &  1 & 1 & 0 & 0 &1 \\[2mm]
&$\op{K}_2$ & 0 &0 &1 &1 &0 &1 \\[1mm] \hline\hline & & & & & & & \\[-3mm]
$\Xi$ & & & & & & & \\
&$\op{K}_1$ & 1 & 0 & 1 & 0 & 1 & 2\\[2mm]
&$\op{K}_2$ & 0 & 0 & 1 & 0 &0 &1\\[1mm] \hline\hline & & & & & & & \\[-3mm]
$V$ & & & & & & & \\
&$\op{K}_1$ & 1 & 1 & 0 & 0 & 1 &1 \\[2mm]
&$\op{K}_2$ & 0 & 1 & 0 & 0 & 0& 1 \\[1mm] \hline\hline 
\end{tabular*}
\end{center}
\end{table}

When the rotating wave approximation is considered, operators of the form
\begin{eqnarray}\label{eq.Kjnl}
\op{K}_\zeta &=&\sum_{s=1}^\ell  \eta_{s}^{(\zeta)} \, \op{\nu}_{s}+ \sum_{k=1}^n \lambda_k^{(\zeta)}\,\op{A}_{kk}\,.
\end{eqnarray}
commute with the Hamiltonian for certain values of the coefficients $\eta_s^{(\zeta)},\,\lambda_k^{(\zeta)}$. These operators $\op{K}_\zeta$ play the role of constants of motion of the system.

For the generalized Dicke Hamiltonian, only the {\it parity} of $\op{K}_\zeta$ is preserved, i.e., the full hamiltonian commutes with operators
\begin{equation}\label{eq.Pij}
\op{\Pi}_\zeta = e^{i\pi\op{K}_\zeta}\,, \quad \zeta=1,\,2\,,\dots\,,\zeta_0\,;
\end{equation}
with $\zeta_0$ denoting the number of parity operators that commute with the Hamiltonian. Its value depends on the particular atomic configuration, and is given in general by~\cite{cordero17}
\begin{equation}
\zeta_0+1=\ell+n -R\,,
\end{equation}  
where $R$ is the rank of the system of algebraic equations
\begin{equation}
\mu_{jk}^{(s)}(\eta_s+\lambda_j -\lambda_k)=0\,,
\end{equation}
in which one needs to take into account all the modes $s=1,2,\dots, \ell$, and the connected pairs for each mode. Notice that when  each mode connects only one pair of levels the number of constants of motion is $\zeta_0=n-1$. On the other hand, when only one mode is responsible for all dipolar transitions one gets $\zeta_0=1$.

Table~\ref{t.K3l} shows the values of the coefficients $\eta_s^{(\zeta)},\,\lambda_k^{(\zeta)}$ for the particular case of $3$-level atoms interacting dipolarly with two modes of electromagnetic field.

The basis in the RWA approximation can be characterized in terms of the eigenvalues $\kappa= \{k_1,\,k_2,\,\dots\,,k_{\zeta_0}\}$ of the constants of motion $\op{K}_\zeta$, and can be written as  
\begin{equation}
{\cal B}_{\textsc{rwa}}^{(\kappa)}:= \left\{ |\vec{\nu};\vec{n}\ket \bigg| \op{K}_\zeta|\vec{\nu};\vec{n}\ket=k_\zeta|\vec{\nu};\vec{n}\ket \textrm{ for } \zeta=1,2\,\dots,\zeta_0\right\}\,. \label{eq.baseRWA}
\end{equation}
For three-level atoms interacting with two modes of electromagnetic field the dimension of this basis can be obtained in analytic form (cf. appendix~\ref{ap.dim}).

When the counterrotating terms are included in the Hamiltonian only the parity of the constants of motion is preserved. 
The Hilbert space then takes into account the direct sum of all the sub-bases ${\cal B}_{\textsc{rwa}}^{(\kappa)}$ for which the parity of each element $k_\zeta$ in $\kappa$ is preserved.

The full basis is then divided in blocks as 
\begin{equation}
{\cal B}=\oplus_\sigma {\cal B}_{\sigma}\,,
\end{equation}
where $\sigma={\rm parity}(\kappa)$, ${\cal B}_\sigma$ is given by
\begin{equation}
{\cal B}_\sigma:= \oplus_{j_1=0}^\infty\cdots \oplus_{j_{\zeta_0}=0}^\infty {\cal B}_{\textsc{rwa}}^{(\kappa_\sigma+2\,\{j_1,j_2,\dots,j_{\zeta_0}\})}\,,
\end{equation}
and $\kappa_\sigma$ is the set of minimum values of the elements of $\kappa$ with the desired parity. The expression ${\kappa_\sigma+2\,\{j_1,j_2,\dots,j_{\zeta_0}\}}$ denotes element by element addition. i.e., $\{k^{{\rm min}}_{1} + 2\, j_1,\dots,k^{{\rm min}}_{\zeta_0}+ 2\,j_{\zeta_0}\}$.

The full Hamiltonian~(\ref{eq.Hnl}) may then be rewritten as $\op{H}=\oplus_\sigma\op{H}_{\sigma}$, and the minimum energy surface is given by
\begin{equation}
{\cal E}_g = \min\{E_{g\sigma}\}\,,
\end{equation}
at each point in parameter space, where $E_{g\sigma}$ is the eigenvalue of $\op{H}_\sigma$ for the ground state:  $\op{H}_\sigma|\Psi_{g\sigma}\ket=E_{g\sigma}|\Psi_{g\sigma}\ket$ for each parity $\sigma$.

\subsection{Truncated Basis via Fidelity}\label{s.truncatednl}

In practice, the exact quantum ground state $|\Psi_{g\sigma}\ket$ is obtained to an approximate precision $|\psi_{g\sigma}^\kappa\ket$ by using the cutoff basis with fixed parity
\begin{equation}\label{eq.Bkappanl}
{\cal B}^\kappa_\sigma:= \oplus_{j_1=0}^{J_1} \oplus_{j_2=0}^{J_2}\cdots \oplus_{j_{\zeta_0}=0}^{J_{\zeta_0}} {\cal B}_{\textsc{rwa}}^{(\kappa_\sigma+2\,\{j_1,j_2,\dots,j_{\zeta_0}\})}\,,
\end{equation}
where $J_i$ is the minimum value of $j_i$ required for  convergence to the desired  precision, of the ground state solution in the Hilbert space.  

In order to calculate the minimum values $\kappa$ in~(\ref{eq.Bkappanl}) which provide a good approximation to the ground state, one may take the variational solution of the problem~\cite{cordero15,cordero17} and propose as minimum values $k_\zeta=\bra\op{K}_\zeta\ket + 3(\Delta K_\zeta)$. This proposal, however, provides a good value $k_\zeta$ only when $\op{K}_\zeta$ obeys a gaussian distribution, and except in the normal region.  Another approach is to use a criterion based on the fidelity between two states.

In this work we use the fidelity criterion to get an approximate ground state. Noticing that the error between the exact and approximate quantum ground state 
\begin{equation}
{\rm e}_{\sigma}^\kappa:= 1-\left|\bra\psi_{g\sigma}^\kappa|\Psi_{g\sigma}\ket\right|^2\,,
\end{equation}
vanishes in the limit $\kappa\to\infty$ and for values $\kappa,\,\kappa+2,\,\dots$, one has ${\rm e}_{\sigma}^\kappa>{\rm e}_{\sigma}^{\kappa+2}>\cdots$, one may cut the full basis to a desirable error ${\rm e}_{\rm rr}$ by imposing the condition 
\begin{equation}\label{criterionl}
1 -F_\sigma^{\kappa} \leq {\rm e}_{\rm rr}\,, \qquad F^\kappa_\sigma := |\bra \psi_{g\sigma}^\kappa|\psi_{g\sigma}^{\kappa+2}\ket|^2\,.
\end{equation} 
This criterion  is more general in that it does not depend on the particular distribution of the values of $\op{K}_\zeta$, and an iterative method allows us to evaluate the value $\kappa$ for the desired approximation.

In order to illustrate the method, we next consider the particular case of a $2$-level system.

\subsubsection{$2$-level system}\label{s.2ls}

We here consider $N_a$ atoms of $2$ levels ($\omega_j<\omega_k$) interacting with a single mode $\Omega_{s}$ of electromagnetic field. The Hamiltonian is  
\begin{eqnarray}\label{eq.H2n}
\op{H}&=&\Omega_{s}\,\op{\nu}_{s} + \omega_j\,\op{A}_{jj} + \omega_k\,\op{A}_{kk} \nonumber \\[2mm]
&-& \frac{1}{\sqrt{N_a}} \mu_{jk}^{(s)} \left(\op{A}_{jk}+\op{A}_{kj} \right)\left(\op{a}^\dag_{s} + \op{a}_{s}\right)\,.
\end{eqnarray}
This possesses only one parity operator, namely
\begin{equation}\label{eq.pi2l}
\op{\Pi}_{jk}^{(s)}=e^{i\pi\op{M}_{jk}^{(s)}}\,, \quad {\rm with}\ \  \op{M}_{jk}^{(s)}=\op{\nu}_{s} + \op{A}_{kk}\,.
\end{equation}
Here $\op{M}_{jk}^{(s)}$ stands for the total number of excitations operator (with eigenvalues $m_{jk}^{(s)}=0,\,1,\,\dots$, if the rotating wave approximation were considered). From the variational calculation one finds that this system presents a phase transition at the critical point 
\begin{displaymath}
\bar{\mu}_{jk}^{(s)} := \frac{1}{2}\,\sqrt{\Omega_{s} \omega_{jk}}\,; \qquad \omega_{jk}:=|\omega_j-\omega_k|\,
\end{displaymath}
which allows us to write the Hamiltonian in terms of the dimensionless dipolar intensity $x_{jk}^{(s)}$ and the detuning parameter $\Delta_{jk}^{(s)}$, given by
\begin{equation}\label{eq.xjk}
x_{jk}^{(s)}:=\frac{\mu_{jk}^{(s)}}{\bar{\mu}_{jk}^{(s)}}\,, \qquad \Delta_{jk}^{(s)}:= \frac{\Omega_{s}}{\omega_{jk}}-1\,,
\end{equation}
so that all the $2$-level systems with the same detuning values $\Delta_{jk}^{(s)}$ have the same quantum phase diagram as function of  $x_{jk}^{(s)}$, i.e., all of these systems are equivalent in these variables.   

\begin{figure}
\includegraphics[width=0.9\linewidth]{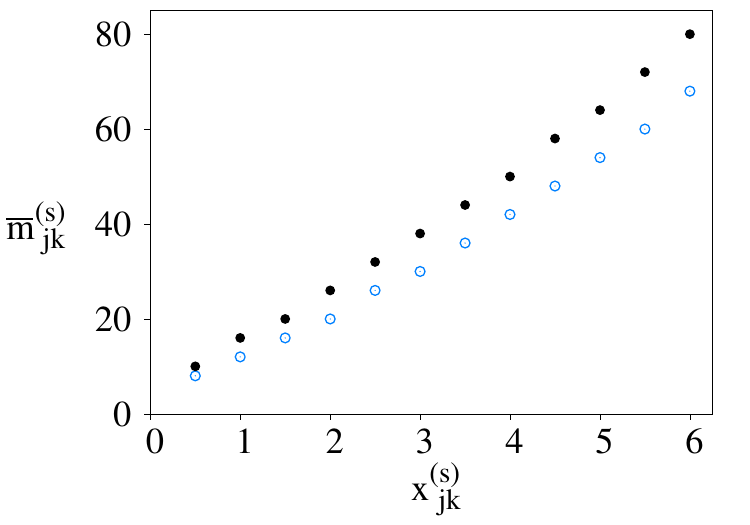}

\caption{For $N_a=3$ atoms with $\Delta_{jk}^{(s)}=0$, the value of $\overline{m}_{jk}^{(s)}$ as a function of $x_{jk}^{(s)}$ is shown for an error of ${\rm e}_{\rm rr}=10^{-10}$ (circles) and ${\rm e}_{\rm rr}=10^{-15}$ (dots) in the fidelity $F$ (Eq.~(\ref{criterionl})).}\label{f.xiMminERR}
\end{figure}

We then calculate, iteratively, a value $\kappa=\{\overline{m}_{jk}^{(s)}\}$ for a fixed parity which will fulfill the inequality in Eq.~(\ref{criterionl}). This value depends of the number of particles $N_a$, the dimensionless dipolar intensity $x_{jk}^{(s)}$, the detuning parameter $\Delta_{jk}^{(s)}$, and the error value ${\rm e}_{\rm rr}$.

In Fig.~\ref{f.xiMminERR}, the value of $\overline{m}_{jk}^{(s)}$ for the even solution $\sigma=e$  is displayed as function of the dimensionless dipolar intensity $x_{jk}^{(s)}$ for the case of $N_a=3$ particles, zero detuning and error values ${\rm e}_{\rm rr}=10^{-10}$ (circles) and ${\rm e}_{\rm rr}=10^{-15}$ (dots). As the error becomes smaller, $\overline{m}_{jk}^{(s)}$ grows considerably, and hence the corresponding dimension of the truncated basis. In fact, it diverges as ${\rm e}_{\rm rr}\to 0$. For the case of the odd parity $\sigma=o$ the number of excitations is given by $\overline{m}_{jk}^{(s)}+1$, with $\overline{m}_{jk}^{(s)}$ the number of excitations for the even solution. 

\begin{figure}
\includegraphics[width=0.9\linewidth]{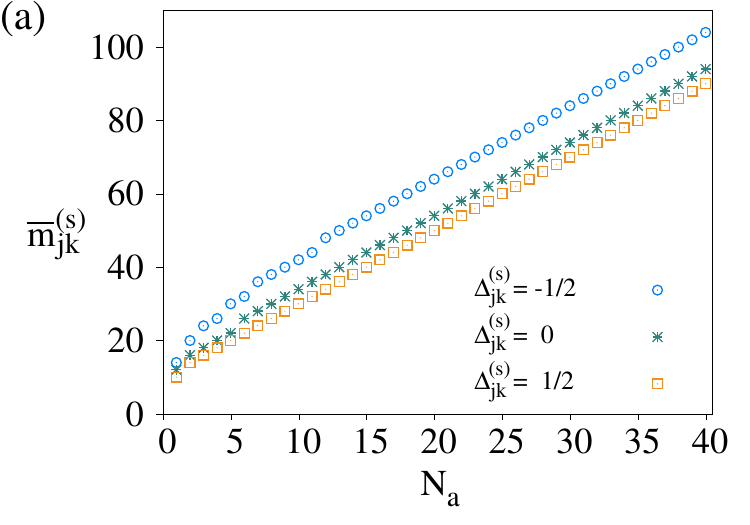}\\
\includegraphics[width=0.9\linewidth]{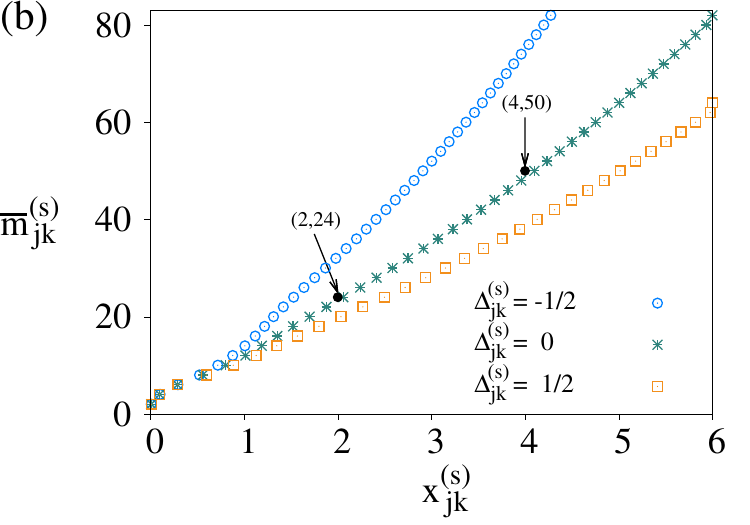}

\caption{Number of excitations $\overline{m}_{jk}^{(s)}$ as a function of: (a) the number of particles $N_a$, with $x_{jk}^{(s)}=3/2$, and (b) the dipolar intensity $x_{jk}^{(s)}$, for $N_a=4$. In both cases  the  detuning values  used are indicated, and we have taken ${\rm e}_{\rm rr}=10^{-10}$.} \label{f.NaMmin}
\end{figure}

Fixing the error value to ${\rm e}_{\rm rr}=10^{-10}$ and taking the even parity $\sigma=e$, Fig.~\ref{f.NaMmin}(a) shows the number of excitations $\overline{m}_{jk}^{(s)}$ as a function of the number of particles $N_a$ for a fixed value $x_{jk}^{(s)}=3/2$ and for different detuning values $\Delta^{(s)}_{jk}=-1/2,\,0,\,1/2$. In a similar way, for $N_a=4$ atoms, the number of excitations are shown as function of $x_{jk}^{(s)}$ in Fig.~\ref{f.NaMmin}(b). The calculation for $\Delta_{jk}^{(s)}=0$ overestimates the values of $\overline{m}_{jk}^{(s)}$ when $\Delta^{(s)}_{jk}>0$. Notice also, that for small values of $x_{jk}^{(s)}$, the $\overline{m}_{jk}^{(s)}$ does not depend on the detuning value.

\subsubsection{General case}

In order to justify the general procedure, it is convenient to first look at a specific case. We take that of $3$-level atoms in the $\Xi$-configuration interacting dipolarly with two modes of an electromagnetic field. The modes $\Omega_1,\, \Omega_2$ promote transitions $\omega_1\rightleftharpoons \omega_2$ and $\omega_2\rightleftharpoons \omega_3$, respectively.  The coefficients of the two operators $\op{K}_\zeta$ Eq.~(\ref{eq.Kjnl}) are given in Table~\ref{t.K3l}, identifying the number of excitations $\op{M}_{jk}^{(s)}$ of each subsystem, and may be written as 
\begin{eqnarray}
\op{K}_1 &=& \op{M}_{12}^{(1)} + \op{M}_{23}^{(2)} + \op{A}_{33}\,, \\[3mm]
\op{K}_2 &=& \op{M}_{23}^{(2)}\,.
\end{eqnarray}

The Hilbert space then divides itself into $4$ subspaces according to the parity of the eigenvalues of $\op{K}_1$ and $\op{K}_2$: $\{ee,\ eo,\ oe,\ oo\}$. For each of these subspaces, the minimum values for $k_1,\ k_2$ that satisfy the convergence criterion ~(\ref{criterionl}) are calculated by using the iterative method described earlier. These values are determined from those of $\overline{m}_{12}^{(1)},\ \overline{m}_{23}^{(2)}$ of the $2$-level subsystems, and from the total number of particles $N_a$. 

Therefore, the maximun eigenvalues that the operators $\op{M}_{12}^{(1)}$ and $\op{M}_{23}^{(2)}$ will take are precisely $\overline{m}_{12}^{(1)}$ and $\overline{m}_{23}^{(2)}$. Using these, and the value of $N_a$ for the operator $\op{A}_{33}$, we arrive at
\begin{eqnarray}
k_1 &=& \overline{m}_{12}^{(1)} + \overline{m}_{23}^{(2)} +N_a\,, \\[3mm]
k_2 &=& \overline{m}_{23}^{(2)}\,.
\end{eqnarray}
Consequently, imposing the condition that the number of excitations of each $2$-level subsystem $m_{jk}^{(s)}$ satisfies the inequality $m_{jk}^{(s)} \leq \overline{m}_{jk}^{(s)} $ for each state of the basis~(\ref{eq.Bkappanl}), the criterion~(\ref{criterionl}) must also be fulfilled. 

For this example, table~\ref{t.dimX} shows the dimension of the basis~(\ref{eq.Bkappanl}) at fixed values of the dimensionless dipolar strength. One may observe how this dimension grows as the number of particles and dipolar strengths increase. 

For the general case of $n$-level atoms interacting with $\ell$-modes we first identify, in the RWA, the number of operators that commute with the Hamiltonian. These are rewritten in terms of the $2$-level subsystem operators $\op{M}_{jk}^{(s)}$ and the weight operators $\op{A}_{kk}$. Their maximum eigenvalues $\overline{m}_{jk}^{(s)}$ are calculated via the iterative procedure described.

For the full Hamiltonian only the parities of the operators $\op{K}_\zeta$ are preserved, and these tell us the number of parity subsystems into which the whole Hilbert space divides. For each one of these parities, we substitute the value of the operators $\op{M}_{jk}^{(s)}$ for $\overline{m}_{jk}^{(s)}$, and the value of the weight operators $\op{A}_{kk}$ for $N_a$. These yield the values for $k_\zeta$ and for all $m_{jk}^{(s)} \leq \overline{m}_{jk}^{(s)} $ the convergence criterion~(\ref{criterionl}) will be fulfilled. 

\begin{table}
\caption{Dimension of the truncated basis ${\cal B}_{ee}^{\kappa}$ Eq.~(\ref{eq.Bkappanl}) for $3$-level atoms in the $\Xi$-configuration interacting with two modes of electromagnetic field, under resonant condition $\Delta_{jk}^{(s)}=0$. The number of allowed photons is restricted by  $\nu_s \leq \overline{m}_{jk}^{(s)} $. This basis permits us to approximate the quantum ground state with a desirable error ${\rm e}_n:=10^{-n}$, at two different values of the dimensionless dipolar strength $(x_{12},x_{23})$ shown.}\label{t.dimX}
\begin{center}
\begin{tabular*}{\linewidth}{c | r | r || r  | r }
 $N_a$ &\phantom{a} ${\rm e}_{10}$  $(1.5,1.5)$\phantom{a} & \phantom{a}${\rm e}_{10}$  $(3,3)$\phantom{a} & ${\rm e}_{15}$ $(1.5,1.5)$ \phantom{a}&\phantom{a}${\rm e}_{15}$ $(3,3)$ \\[1mm] \hline\hline & & &\\[-2mm]
1 & 91 \phantom{aaa}& 271 \phantom{a}&  169 \phantom{aaa}& 397 \phantom{a} \\
2 & 330 \phantom{aaa}& 925 \phantom{a}& 532  \phantom{aaa}& 1\,426 \phantom{a}\\
3 & 664 \phantom{aaa}& 2\,295 \phantom{a}& 1\,030  \phantom{aaa}& 3\,667 \phantom{a}\\
4 & 1\,222 \phantom{aaa}& 4\,876 \phantom{a}&  2\,170 \phantom{aaa}& 7\,956 \phantom{a} \\
5 & 2\,017 \phantom{aaa}& 9\,090 \phantom{a}& 3\,442  \phantom{aaa}& 13\,985 \phantom{a}\\[1mm] \hline
\end{tabular*}
\end{center}
\end{table}
%%%

A good  estimate for the dimension of the basis is given by
\begin{equation}
\frac{1}{\zeta_0} \prod_{s=1}^\ell (\overline{m}_{jk}^{(s)} +1) \binom{N_a+n-1}{n-1}\,,
\end{equation}
where $\zeta_0$ is the number of parities preserved, which in our example of table~\ref{t.dimX} is $\zeta_0=4$, together with $\ell=2$ and $n=3$.

\subsection{Minimum Energy Surface}\label{s.DTCenergia}

The minimum energy surface of physical systems lets us determine the quantum phase transitions at zero temperature~\cite{kittel65}.  In the Dicke model the quantum phase transitions were determined by Hepp and Lieb~\cite{hepp73} and the free energy of the system in the thermodynamic limit calculated by Wang and Hioe~\cite{wang73}.  A review of the dynamics of matter-field interactions of two- and three-level systems was done by Eberly and Yoo~\cite{yoo85}. 
A procedure to determine the quantum phase transitions was proposed by Gilmore~\cite{gilmore93}, which uses a variational test function together with the catastrophe formalism. Another possibility to determine the quantum phase transitions (also called crossovers) for a finite number of particles is by means of the fidelity concept of quantum information~\cite{jozsa94, zanardi06, cordero19, castanos18b}. Here we illustrate how to build the ground state energy surface of $N_a=4$ atoms of $3$ levels interacting dipolarly with two modes of an electromagnetic field, together with the determination of the quantum phase diagram through the calculation of the fidelity. 

We choose the parameters in the Hamiltonian to be: atomic levels $\omega_1=0,\, \omega_2=1/4$ and $\omega_3=1$; field frequencies $\Omega_1=1/4$ and $\Omega_2=3/4$; and as phase space parameters the dimensionless dipolar strengths $x_{12}^{(1)}$ and $x_{23}^{(2)}$. The superscripts indicate that the modes $\Omega_1$ and $\Omega_2$ promote transitions between the atomic levels $\omega_1\rightleftharpoons \omega_2$ and $\omega_2\rightleftharpoons \omega_3$, respectively. Note that the system is in double resonance, i.e., the case of zero detuning.
\begin{figure}
\begin{center}
\includegraphics[width=0.8\linewidth]{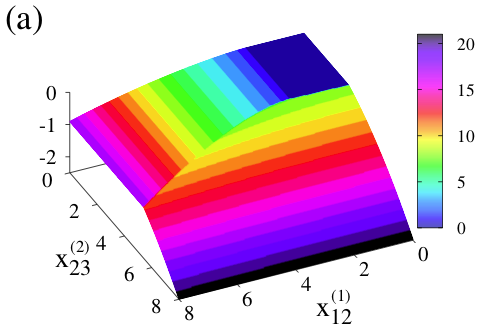}\\
\includegraphics[width=0.8\linewidth]{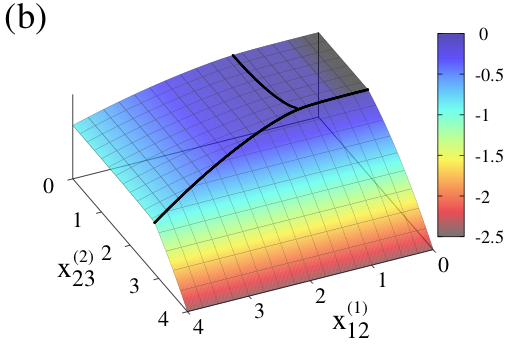}
\end{center}
\caption{Color online. Quantum ground energy surface for $N_a=4$ atoms in the $\Xi$-configuration. (a) Generalized Tavis-Cumming model, where $k_1$ is a constant; the color indicates the value of the total number of excitations in the system. (b) Generalized Dicke model, where only the parity of $k_1$ is conserved; the black lines define the separatrix where a minimum in the fidelity occurs. Parameters used are discussed in the text}\label{f.eminX}
\end{figure}

Recall that the basis ${\cal B}_{\textsc{rwa}}^{(\kappa)}$ in Eq.~(\ref{eq.baseRWA}) allows us to calculate the ground state energy surface of the generalized Tavis-Cumming model where the operators $\op{K}_\zeta$ are constants of motion. In this case one can evaluate the ground state energy surfaces for fixed values of $\kappa$, and then take the minimum value as a function of the control parameters. On the other hand, the minimum energy surface for the generalized Dicke model requires the use of the basis ${\cal B}_{\sigma}^{\kappa}$ in Eq.~(\ref{eq.Bkappanl}). One then evaluates the minimum energy surfaces for every fixed parity of $\kappa$, and takes the one which is minimum at every point in phase space. For atoms in the $\Xi$-configuration the ground state energy surface has even-even parity $\sigma=ee$ when the number of particles is even, while it is composed of the parities $\sigma=ee$ and $\sigma=oe$ for an odd number of particles (depending on the region of phase space). A similar situation occurs for the one atom $\Lambda$-configuration~\cite{cordero19}. 

In the Tavis-Cumming model we have $2$ constants of motion, viz., the total number of excitations $k_1$, and the number of excitations of the $2 \rightleftharpoons 3$ subsystem $k_2$. Its energy surface is plotted in Fig.~\ref{f.eminX}(a) for $N_a=4$, and we see that the phase diagram is divided into $3$ regions: a normal region where ${\cal E}_g = 0$, and a collective region showing a separatrix between its two subregions. The height indicates the energy value (in units of $\hbar \omega_3$) and the color gives the value of $k_1$; in this approximation $k_2=0$ for the smaller values of $x_{23}$ where there are only photons of mode $\Omega_1$.  For the collective region where there are only photons of mode $\Omega_2$,  $k_2$ increases gradually from $3$ in steps of the unity. Note that all the transitions are discontinuous because the ground state changes from one subspace to another as $k_1$ changes. The transitions where $k_1$ is increased (or decreased) by a value greater than one remain as discontinuous transitions in the thermodynamic limit. When leaving the normal region as $x^{(1)}_{12}$ increases the transition is continuous; when leaving it as   $x^{(2)}_{23}$ increases, the transition is discontinuous.

For the generalized Dicke model $k_1$ and $k_2$ are no longer constants of motion, only their parity is preserved and they distribute about their corresponding GTCM-values, taking into account that the dimensionless coupling strengths $x_{ij}$ scale by a factor of $1/2$. The ground state energy surface is plotted in Fig.~\ref{f.eminX}(b). Since $N_a$ is even the ground state energy surface has an even-even parity. We also show the separatrix (black points), obtained from the local minima in the fidelity between neighbouring points. Here one finds a second order transition from the normal region (the enclosed region around the origin) to the collective region that we reach by increasing the value of $x_{12}^{(1)}$; all other transitions are first order discontinuous transitions.

\section{Reduced Bases}\label{s.redB}

We have shown how the dimension of the truncated basis ${\cal B}_\sigma^\kappa$ Eq.~(\ref{eq.Bkappanl}) grows quickly as both the number of particles and the dimensionless parameter control $x_{jk}^{(s)}$ increase.  However, for a fixed desirable error ${\rm e}_{\rm rr}$ (as for example ${\rm e}_{\rm rr}=10^{-10}$), it is clear that any value less than ${\rm e}_{\rm rr}$ in the calculation is negligible. Thus, in principle one may discard all the weakly coupled  states in the ground state with a joint probability less than ${\rm e}_{\rm rr}$, defining in this way a reduced basis and obtaining a good approximation for the quantum ground state. 

The variational solution is used as a criterion that allows us to discard weakly coupled states of the basis. The variational solution of this kind of systems shows that the collective region is divided into sub-regions where only one kind of photon contributes to the ground state, while the other ones remain in the vacuum state.  In fact, in each subregion the full system has a behavior similar to a subsystem with a single mode \cite{cordero15,cordero16,cordero17} except in a small vicinity of the separatrix. This behavior was exhibited for the case of a single particle~\cite{cordero19}. 
 
In order to discard states in the full basis, we consider the two sectors, matter and field, separately.

\subsection{Matter Sector}

The procedure to extract the significant states from the matter sector is based on the following statement: 
By considering the case where a single mode $\Omega_s$ promotes the transitions between a pair of atomic levels, that is, $x_{jk}^{(s)} \neq 0$, we have $x_{jk}^{(s')} = 0$ for $s'\neq s$.  In this case, the variational solution shows that the collective region divides itself into $\ell_0$ sub-regions (here $\ell_0=\ell$, but in general $\ell_0\geq\ell$) where, in each of them, a two level Hamiltonian of the form given in  Eq.~(\ref{eq.H2n}) dominates~\cite{cordero15}.  For the variational solution one finds that the parity of the operator $\op{M}_{jk}^{(s)}$ in Eq.~(\ref{eq.pi2l}) is preserved and also $\op{A}_{jj} + \op{A}_{kk}=N_a\,\op{1}_{matt}$ is fulfilled. Therefore we define the number of particles of each subsystem as 
\begin{equation}
N_{jk}^{(s)} := \bra a_1,\,\dots,\,a_n | \op{A}_{jj}+\op{A}_{kk} \,|a_1,\,\dots,\,a_n\ket \,.
\label{Njk}
\end{equation}
Since the variational solution of the matter sector has contributions of states where at least one  $N_{jk}^{(s)}$ takes the value $N_a$, one may classify the matter subbasis by the set of states ${\cal M}_r$ where  $N_{jk}^{(s)} = N_a-r$ is satisfied at least for one subsystem, i.e. 
\begin{equation}\label{eq.Mrnl}
{\cal M}_r := \left\{ |a_1,\,\dots,\,a_n\ket \big| \vee_s N_{jk}^{(s)} = N_a-r \right\}\,,
\end{equation}
where $\vee_s$ is the logical ``or'' operator.
Notice that ${\cal M}_0$ is the matter contribution according to the variational solution and hence this contribution remains in the thermodynamic limit, while the other ones (for $r>0$) vanish as $N_a$ grows. Also note that the full matter basis~(\ref{eq.Mnl}) is given by 
\begin{equation}\label{eq.Mrsumnl}
{\cal M} = \oplus_{r=0}^{O_1} {\cal M}_r\,,
\end{equation}
where $O_1$ is the maximum number of matter subbases and it is given in general by the floor of (greatest integer less than or equal to) $N_a/ \ell_0$
\begin{equation}\label{eq.o1}
O_1=\left\lfloor \frac{N_a}{\ell_0} \right\rfloor\,,
\end{equation}
relationship that is obtained from the inequalities \[N_a\leq N_{jk}^{(1)}+\cdots+N_{j'k'}^{(\ell_0)} \leq \ell_0\,N_a\,.\]

From expression~(\ref{eq.Mrsumnl}) one may consider different orders to the matter contribution 
\begin{equation}\label{eq.Mo1nl}
{\cal M}[o_1] := \oplus_{r=0}^{o_1} {\cal M}_r\,,\qquad o_1\leq O_1.
\end{equation}

As an example of the division of the matter sector, consider four $3$-level atoms in the $\Xi$-configuration interacting dipolarly with two modes of electromagnetic field. In this case we have two $2$-level subsystems, as described above, $N^{(1)}_{12} = a_1+a_2$, $N^{(2)}_{23} = a_2+a_3$, and
$$
{\cal M}_r := \left\{ |a_1,\,a_2,\,a_3\ket \big| a_1+a_2=4-r \vee a_2+a_3=4-r \right\}\,,
$$
which gives rise to
\begin{center}
\begin{tabular}{c | c }
sub-basis & states \\ \hline &\\[-3mm]
${\cal M}_0$ & $\vert a_1\, a_2\, 0\rangle$; $\vert 0\, a_2\, a_3\rangle$  \\ [1mm]
 ${\cal M}_1$ & $\vert a_1\, a_2\, 1\rangle$; $\vert 1\, a_2\, a_3\rangle$  \\ [1mm]
 ${\cal M}_2$ & $\vert a_1\, a_2\, 2\rangle$; $\vert 2\, a_2\, a_3\rangle$ 
\end{tabular}
\end{center}
Here, $a_1+a_2+a_3=4$, the number of particles.

\begin{figure*}
\begin{center}
%B0
Subset ${\cal M}_0$\\[2mm]
\includegraphics[width=0.11\linewidth]{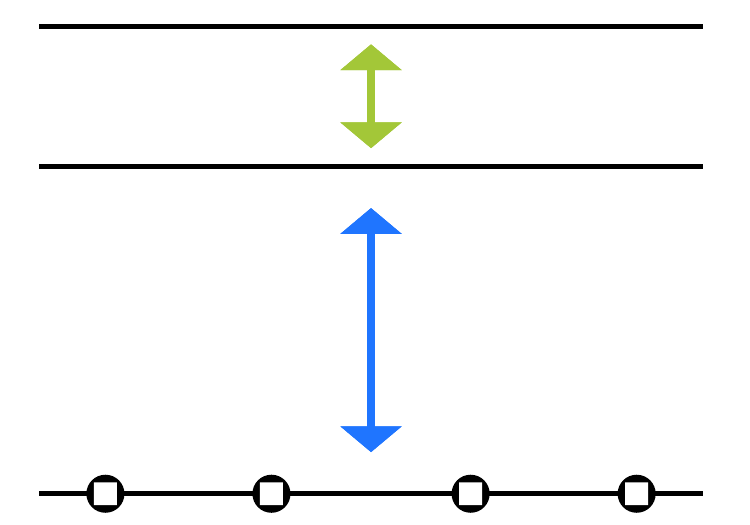}\hspace{2mm}
\includegraphics[width=0.11\linewidth]{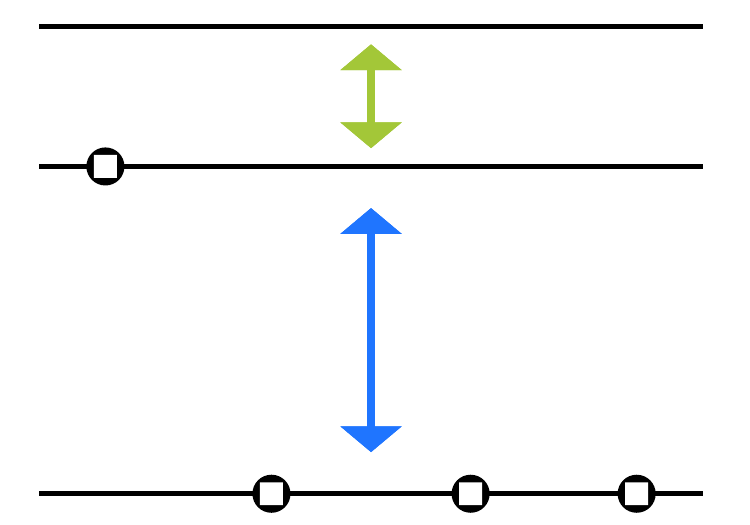}\hspace{2mm}
\includegraphics[width=0.11\linewidth]{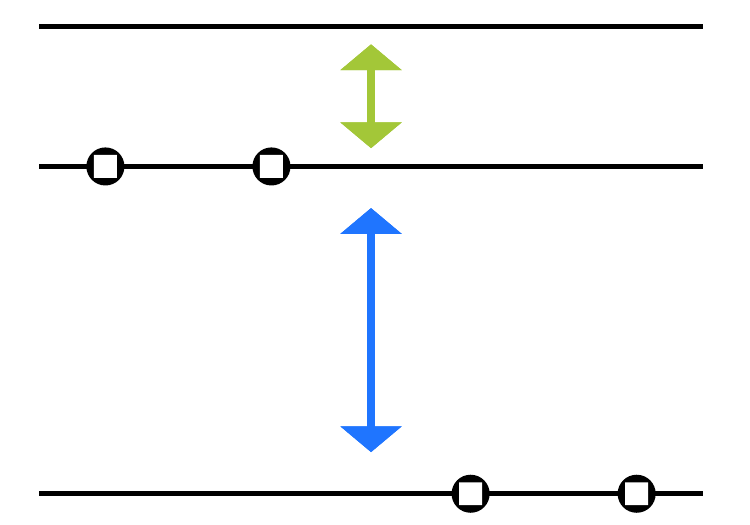}\hspace{2mm}
\includegraphics[width=0.11\linewidth]{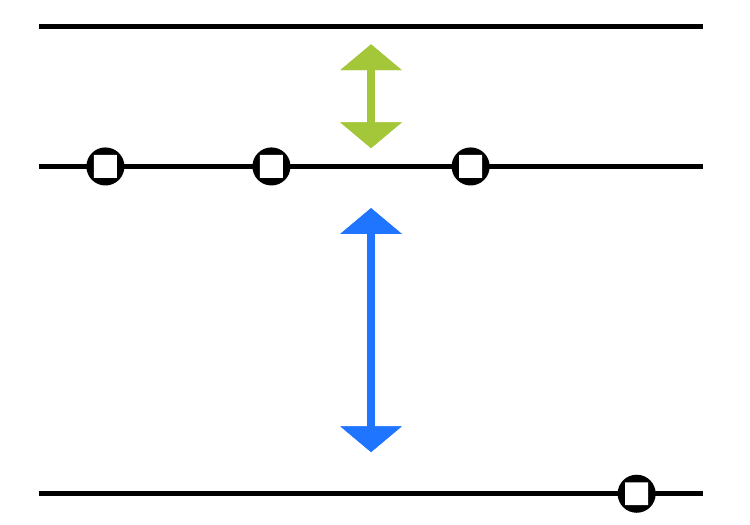} \\[5mm] 
\includegraphics[width=0.11\linewidth]{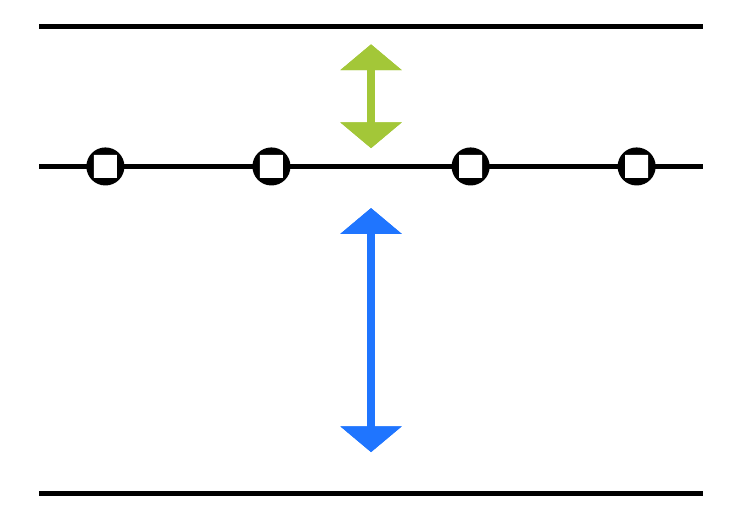}\hspace{2mm}
\includegraphics[width=0.11\linewidth]{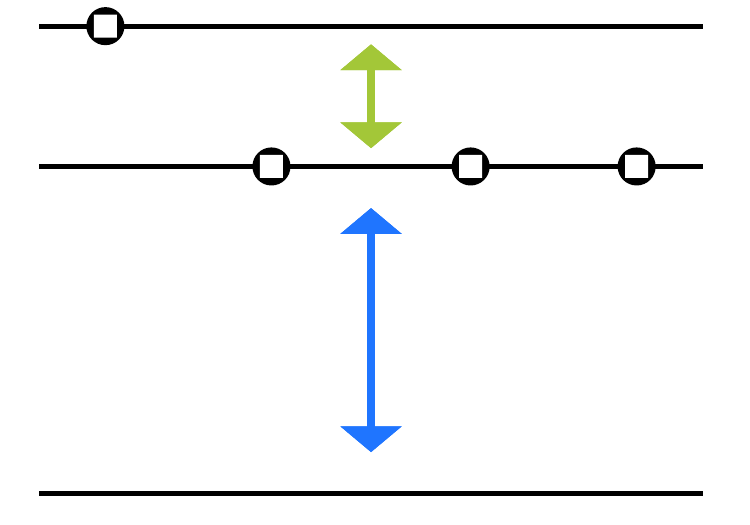}\hspace{2mm}
\includegraphics[width=0.11\linewidth]{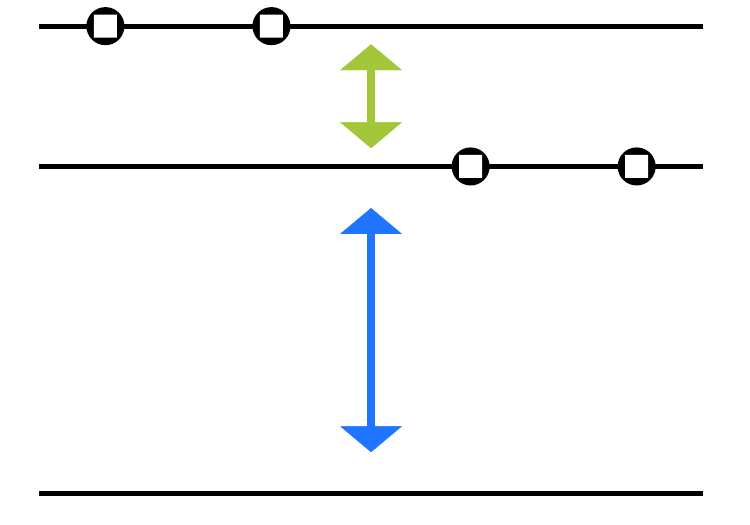}\hspace{2mm}
\includegraphics[width=0.11\linewidth]{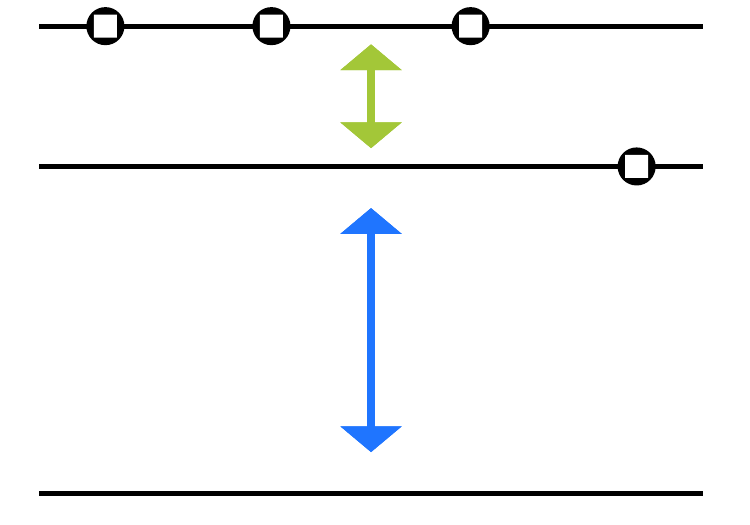}\hspace{2mm}
\includegraphics[width=0.11\linewidth]{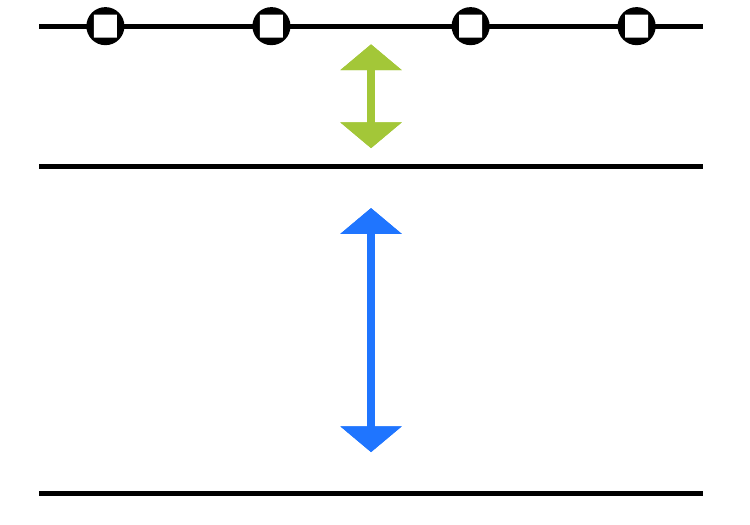}\\[10mm]
%B1
Subset ${\cal M}_1$\\[2mm]
\includegraphics[width=0.11\linewidth]{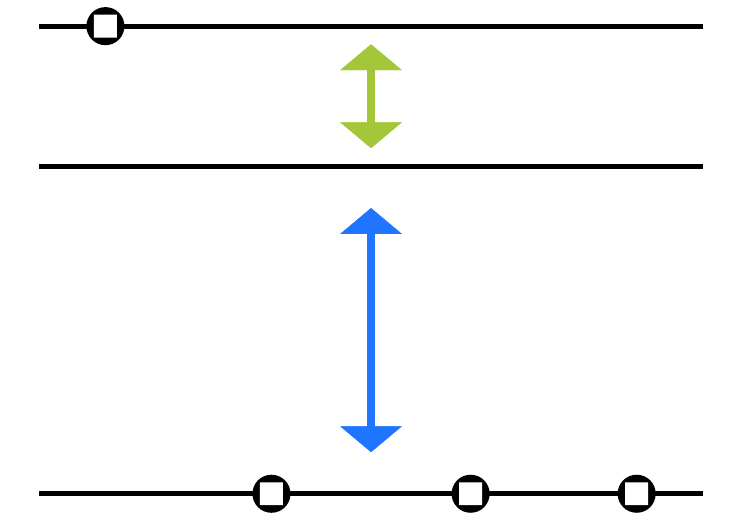}\hspace{2mm}
\includegraphics[width=0.11\linewidth]{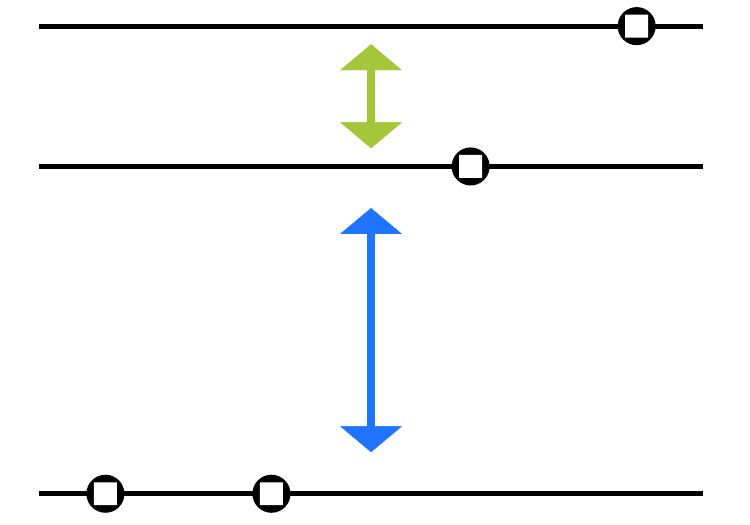}\hspace{2mm}
\includegraphics[width=0.11\linewidth]{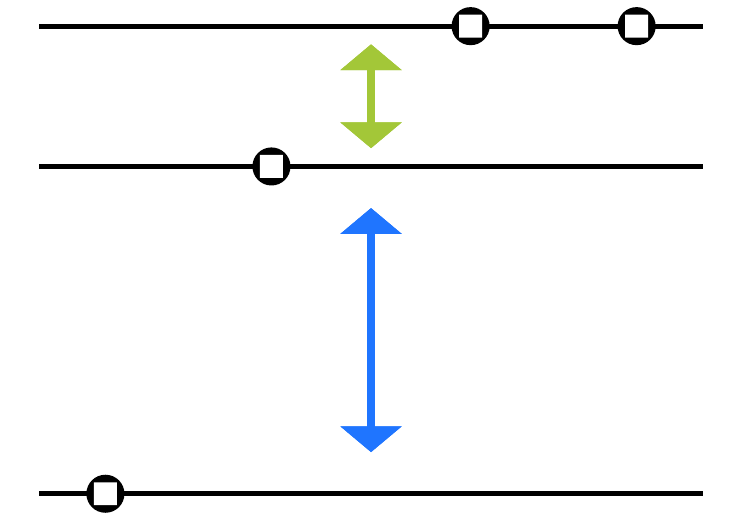}\hspace{2mm}
\includegraphics[width=0.11\linewidth]{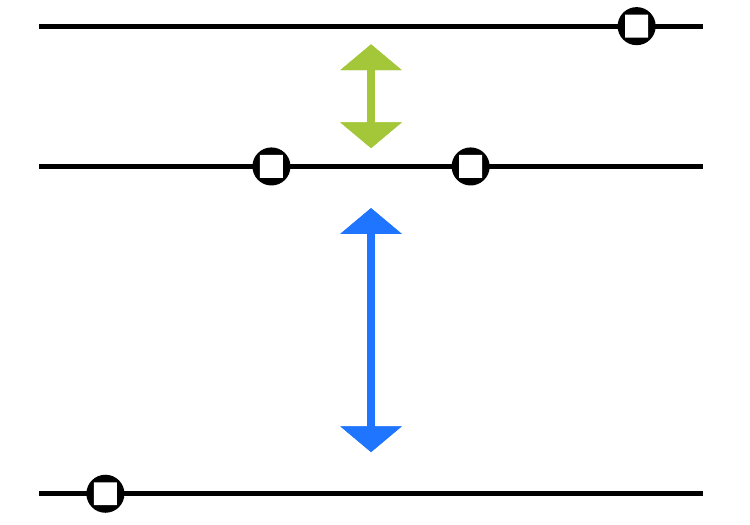}\hspace{2mm} 
\includegraphics[width=0.11\linewidth]{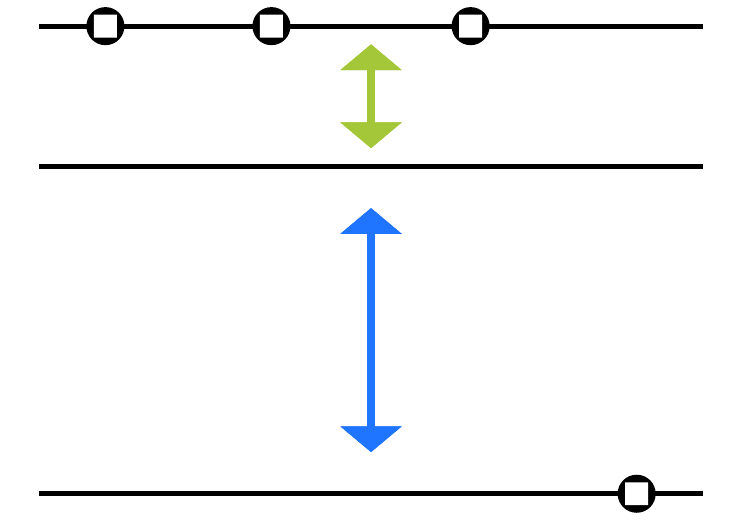} \\[10mm] 
%B2
Subset ${\cal M}_2$\\[2mm]
\includegraphics[width=0.11\linewidth]{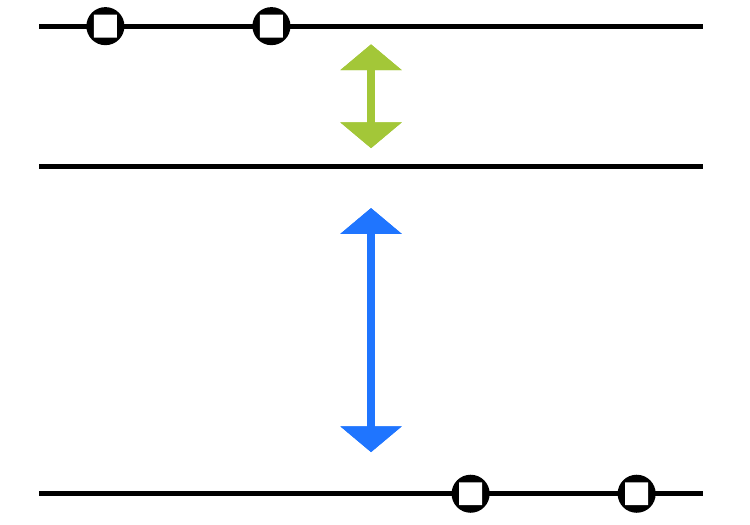} 
\end{center}
\caption{For $N_a=4$ atoms in the $\Xi$-configuration, the matter states are shown for each defined matter subspace ${\cal M}_r$. Horizontal lines denote atomic levels, and circles denote atomic occupations.}\label{fXNa4}
\end{figure*}

 In Fig.~\ref{fXNa4} we show schematically the states in each ${\cal M}_r$. The variational solution for the ground state has matter sector ${\cal M}_0$~\cite{cordero15,cordero17}.  Hence, ${\cal M}[o_1]$ with $o_1=1,\,2$ provides the corrections in the matter sector due to the entanglement  between the subsystems.

In the general case, when a single mode may promote transitions of two or more atomic level pairs, Eqs.~(\ref{eq.Mrnl})-(\ref{eq.Mo1nl}) are the same but the expression of $N_{jk}^{(s)}$ is replaced by  $N^{(s)}$, which takes the form of the sum of the matter weight operators $\op{A}_{kk}$  that describe the $s$-th subsystem. One may find the different subsystems by breaking down the full atomic configuration in parts where only one mode connects the atomic levels as was shown in~\cite{cordero16,cordero17}.

\subsection{Field Sector}

In the truncated basis of ${\cal B}_\sigma^\kappa$ Eq.~(\ref{eq.Bkappanl}), the maximum photon contribution $\tilde{\nu}_{s}$ for each mode $\Omega_{s}$ depends on the value of $\kappa$. Since in general there is no simple relationship between $\kappa$ and $\tilde{\nu}_{s}$, we take without loss of generality $\tilde{\nu}_{s}:=\nu_0=\max(\kappa)$. In the end we eliminate the states that do not satisfy the parity and upper limits of the truncated basis. So, for the value  $\nu_0$,  the truncated Fock basis of each mode is
\begin{equation}
{\cal F}_{s}[\nu_0] := \left\{ |\nu_{s}\ket \big| \nu_{s} \leq \nu_0\right\}\,,
\end{equation}
and the truncated field sector is thereby given in general as
\begin{equation}
{\cal F} = \otimes_{s=1}^\ell {\cal F}_{s}[\nu_0]\,.
\end{equation}

We next want to sub-divide the different photon contributions. Since the Hamiltonian interaction~(\ref{eq.Hnl}) connects  the state $|\nu_{s}\ket$ with the states $|\nu_{s}+1\ket$ and $|\nu_{s}-1\ket$, and using the fact that in the variational solution for each subsystem $\op{H}_{s}$ the contribution of the mode $\Omega_{s'}$ ($s'\neq s$) is negligible, one may truncate the contribution of negligible photons by taking
\begin{equation}
{\cal F}[o_2] = \oplus_{s=1}^\ell \left[\otimes_{s'}^\ell {\cal F}_{s'}[\zeta_{ss'}]\right]\,, 
\end{equation}
with
\begin{equation}
\zeta_{ss'}=\left\{\begin{array}{l l} \nu_0 & s=s' \\[2mm] 2\,o_2+1 & s\neq s' \end{array} \right.\,.
\end{equation}
Here, $o_2$ is the order in the field sector, which can take the maximum value
\begin{equation}\label{eq.o2}
O_2 :=\left\lfloor \frac{\nu_0}{2} \right\rfloor\,.
\end{equation}

As an example of how to truncate the field sector, we consider as before $4$ atoms in the $\Xi$ configuration with two photon modes, one for each two-level subsystem. 
By considering $x_{12}^{(1)}=2 $ and $x_{23}^{(2)}=4$, one determines (see Fig.~\ref{f.NaMmin}) 
\begin{equation*}
\overline{m}^{(1)}_{12} = 24    \, , \quad \overline{m}^{(2)}_{23} =  50\, .
\end{equation*}
Therefore the minimum values for the constants of motion to achieve convergence to the required value of ${\rm e}_{\rm rr}=10^{-10}$ are given by
\begin{equation*}
k_1= \overline{m}^{(1)}_{12} + \overline{m}^{(2)}_{23} +4=78   \, , \quad k_2 = \overline{m}^{(2)}_{23} = 50 \, . 
\end{equation*}

The basis states are given by
\[
{\cal F}[o_2]  = {\cal F}_1[\nu_0] \otimes {\cal F}_2[2 \, o_2+1] \oplus {\cal F}_1[2 \, o_2+1] \otimes {\cal F}_2[\nu_0]  \, ,
\]
where $\nu_0=78$,  $0\leq o_2 \leq 39$ and the dimension is given by 
\begin{eqnarray}
{\rm Dim}(F[o_2])&=&4\,(\nu_0+1)(o_2+1) -  4( o_2+1)^2 \nonumber \\[2mm]
&=&4\,(\nu_0-o_2)( o_2+1)\,,
\end{eqnarray}
in comparison with the dimension of the full field basis given by ${\rm Dim}[{\cal F}]=( \nu_0+1)^2$. Notice  the difference in the cardinality of the different bases: 
\begin{center}
\begin{tabular}{c | c }
sub-basis & Dim \\ \hline &\\[-3mm]
${\cal F}[0] $ &  \phantom{6\,}312  \\ [1mm]
 ${\cal F}[1] $ &  \phantom{6\,}616 \\ [1mm]
  ${\cal F}[2] $ &  \phantom{6\,}912 \\ [1mm]
 ${\cal F}\phantom{[2]}$ &   6\,241
\end{tabular}
\end{center}

\subsection{Matter-Field Sector}

For the combined matter-field system, using the definition of the truncated basis in the matter and field sectors, we take the reduced basis to be
\begin{equation}\label{eq.Bro1o2nl}
{\cal B}_\sigma^\kappa[o_1,o_2]:= \Big[ \oplus_{r_1=0}^{o_1}\oplus_{r_2=0}^{o_2}{\cal F}[r_2] \otimes {\cal M}[r_1] \Big]_\sigma^\kappa \,.
\end{equation}
Indices  $\kappa$ and $\sigma$ indicate that  states that do not preserve the parity $\sigma$ and states with values $\kappa'>\kappa$ are eliminated. Notice that one has the identity 
\begin{equation}
{\cal B}_{\sigma}^\kappa = {\cal B}_\sigma^\kappa\left[O_1,O_2\right]\,.
\end{equation}
We should remark that this procedure to obtain reduced bases is useful only when the full system is divided into subsystems where a single mode promotes transitions between a few atomic levels. In addition, for large values of $N_a$ the reduction ${\cal B}_\sigma^\kappa[o_1,o_2]$ will give a good approximation to the exact quantum ground state, because this state approaches better and better the symmetry-adapted variational case.

\section{Results for the $\Xi$-Configuration}\label{s.Xiconf}

As an example of the application of the reduced basis, we  consider a $3$-level system in the $\Xi$-configuration in resonance with two modes of electromagnetic field (zero detuning).

\subsection{Dimensions for Different Orders}\label{s.dimensions}

In order to compare the dimension of the reduced basis~(\ref{eq.Bro1o2nl}) with the exact basis~(\ref{eq.Bkappanl}), we fix the dimensionless dipolar strength values at $x_{12}^{(1)}=x_{23}^{(2)}=4$ and the error in the fidelity at ${\rm e}_{\rm rr}=10^{-10}$. For these equal values of $x_{12}^{(1)}$ and $x_{23}^{(2)}$ the values for $\overline{m}_{12}^{(1)}$ and $\overline{m}_{23}^{(2)}$ are equal. The bases will allow us to find the ground state as function of the parameters in the region $[0,4]\times[0,4]$ as in Fig.~\ref{f.eminX}(b).

By definition of the reduced bases, these satisfy 
\begin{eqnarray} 
{\cal B}_\sigma^\kappa[o_1,o_2] &=& {\cal B}_\sigma^\kappa[O_1,o_2]\,,\qquad \textrm{when} \ o_1\geq O_1\,, \nonumber\\
{\cal B}_\sigma^\kappa[o_1,o_2] &=& {\cal B}_\sigma^\kappa[o_1,O_2]\,,\qquad \textrm{when} \ o_2\geq O_2\,.\nonumber
\end{eqnarray}
In this sense one may refer to any order of approximation independently of the number of particles $N_a$, which establishes the value $O_1$ in Eq.~(\ref{eq.o1}), and maximum number of photons $\nu_{0}$ for the value of $O_2$ in Eq.~(\ref{eq.o2}). 
\begin{figure}
\includegraphics[width=0.9\linewidth]{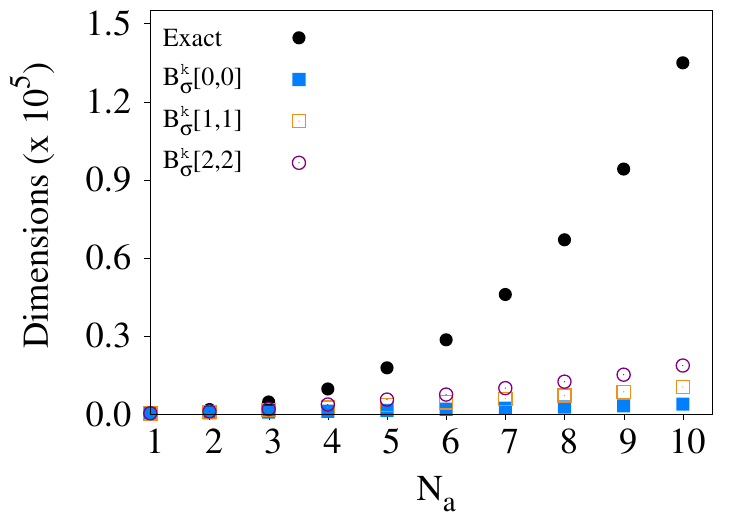}
\caption{The dimension of the Hilbert space~(\ref{eq.Bkappanl}) is shown as a function of the number of particles $N_a$, considering an error ${\rm e}_{\rm rr}=10^{-10}$ in the fidelity  and maximum values of $x_{12}=x_{23}=4$, for the $\Xi$-configuration (solid circles) and even-even parity $\sigma=ee$. This is compared to the corresponding  dimensions of the reduced bases~(\ref{eq.Bro1o2nl}) with orders $o1=o2=0$ (solid squares), $o1=o2=1$ (empty squares)  and $o1=o2=2$ (empty circles) .}\label{f.dimQasNa}
\end{figure}

The dimensions of the reduced bases as function of the number of particles are shown in Fig.~\ref{f.dimQasNa} for the even-even parity ($\sigma=ee$). Note that the savings are tremendous. In particular, for $N_a=1$ and $N_a=10$ we have
\begin{center}
\begin{tabular*}{\linewidth}{c | c | c}
\phantom{aa} Basis \phantom{a} &  \phantom{aa} dim. for $N_a=1$ \phantom{a} & \phantom{aa} dim. for $N_a=10$ \\ \hline\hline &&\\[-3mm]
${\cal B}_\sigma^\kappa$ & 397 &133\,549\\[1mm]
${\cal B}_\sigma^\kappa[2,2]$ & 252 & \phantom{1}18\,452 \\[1mm]
${\cal B}_\sigma^\kappa[1,1]$ &216& \phantom{1}10\,226 \\[1mm]
${\cal B}_\sigma^\kappa[0,0] $ &  176 & \phantom{13}3\,754 \\[1mm]
\end{tabular*}
\end{center}

In order to calculate the table above, we used 
$$
\overline{m}_{12}^{(1)} = \overline{m}_{23}^{(2)} = 22 \quad {\rm for}  \quad N_a=1,
$$
and 
$$
\overline{m}_{12}^{(1)} = \overline{m}_{23}^{(2)} = 92 \quad  {\rm for} \quad N_a=10,
$$
which imply that $(k_1,\,k_2) = (45, 22)$ and $(194, 92)$ respectively. Clearly the reduced bases will be more important in calculations where the number of particles is large.

\subsection{Comparison between energy surfaces}

Previously, we have shown the exact ground energy surface for the case $N_a=4$ in Fig.~\ref{f.eminX}(b). For this case we find that the dimensions of the reduced bases are: ${\rm dim}\,({\cal B}_\sigma^\kappa[0,0])=1020$, ${\rm dim}\,({\cal B}_\sigma^\kappa[1,1])=2413$ and ${\rm dim}\,({\cal B}_\sigma^\kappa[2,2])=3609$, in comparison with the dimension of the exact quantum basis  ${\rm dim}\,({\cal B}_\sigma^\kappa)=9546$. 

To compare the different energy surfaces, we define $E_{o1,o2}$ as the ground state energy by using the reduced basis ${\cal B}_\sigma^\kappa[o_1,o_2]$ and calculate the percentual error with respect to the exact quantum ground energy ${\cal E}_g$ as
\begin{equation}
\Delta_{o1,o2}= \left|\frac{({\cal E}_g-E_{o1,o2})}{{\cal E}_g}\right|\,.
\end{equation}
We define $\Delta_{o1,o2}=0$ when ${\cal E}_g=0$, since this value is obtained at  points on the axes and one may see easily that any basis reduction provides the same results as the exact basis when ${\cal E}_g=0$.
\begin{figure}
\begin{center}
\includegraphics[width=0.8\linewidth]{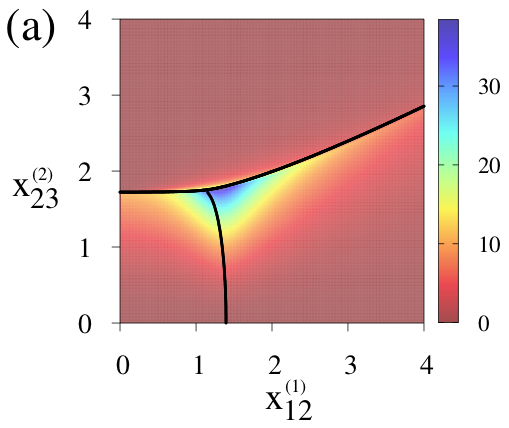}\\
\includegraphics[width=0.8\linewidth]{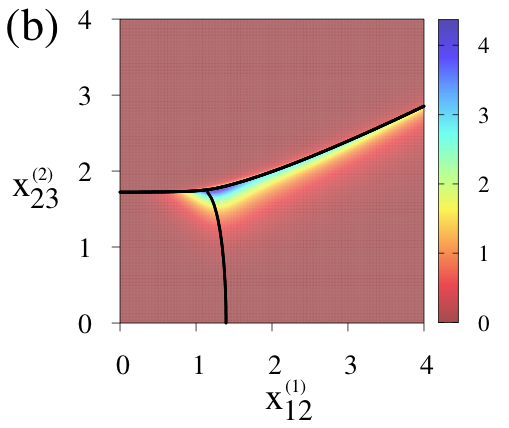}\\
\includegraphics[width=0.8\linewidth]{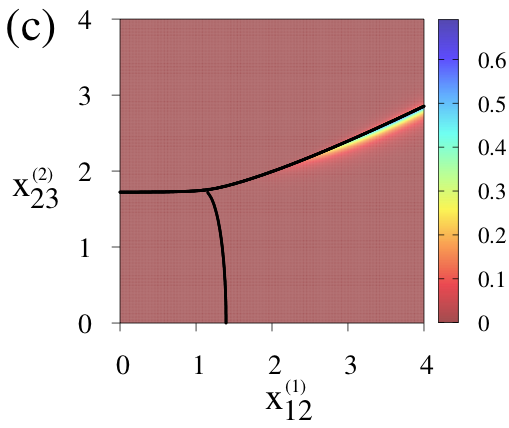}
\end{center}
\caption{Percentual error $\Delta_{o1,o2}$ in the quantum ground energy surface for the reductions: (a) ${\cal B}_\sigma^\kappa[0,0]$, (b)  ${\cal B}_\sigma^\kappa[1,1]$ and (c)  ${\cal B}_\sigma^\kappa[2,2]$.  Note that the plots are given at different scales. The parameters are discussed in the text.}
\label{f.error}
\end{figure}

We choose the parameters for the system indicated in section~\ref{s.DTCenergia} and compare the energy surfaces in Fig.~\ref{f.error} for the reductions ${\cal B}_\sigma^\kappa[0,0]$,  ${\cal B}_\sigma^\kappa[1,1]$, and ${\cal B}_\sigma^\kappa[2,2]$. In all cases, one may observe that the maximum percentual error is around the separatrix, where discontinuos transitions occur, while away from the separatrix the percentual error tends rapidly to zero; that is, we find an excellent agreement with the exact value for the ground state energy when using our reduced bases.

\subsection{Photon Number Fluctuations}
\begin{figure}
\begin{center}
\includegraphics[width=0.8\linewidth]{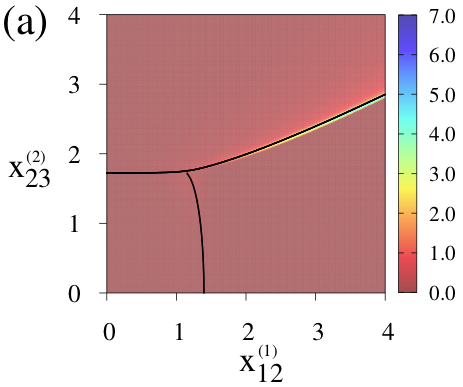}\\
\includegraphics[width=0.8\linewidth]{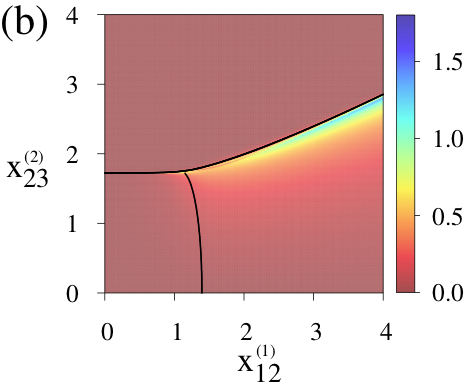}
\end{center}
\caption{Absolute error in the fluctuations of the number of photons $\Delta(\sigma_\nu)$ for the quantum ground state in comparison with the ground state in the reduction ${\cal B}_\sigma^\kappa[0,0]$, (a) for photon $\Omega_1$, and (b) for photon $\Omega_2$. Note that the plots are given at different scales.}
\label{f.flucerror}
\end{figure}

In the previus subsection we saw that the value of the ground state energy found with the reduced bases is in very good agreement  with the exact quantum calculation. Similar results are obtained for any expectation value of both number of photons or atomic populations. Here, we show the absolute error in the fluctuations in the number of photons
\begin{equation}
\Delta(\sigma_\nu):= |\sigma_g(\nu)-\sigma_{o1,o2}(\nu)|\,.
\end{equation}
a quantity that is not well approximated by variational methods.

In Fig.~\ref{f.flucerror} the fluctuation errors are shown for the calculations with the reduced basis ${\cal B}_\sigma^\kappa[0,0]$, for photons $\Omega_1$ in Fig.~\ref{f.flucerror}(a) and for photons $\Omega_2$ in Fig.~\ref{f.flucerror}(b). Again the maximum error is around the separatrix. This error vanishes when the order of the reduced basis increases. This shows that in addition to the expectation values one has an excellent agreement for their fluctuations, and hence the ground state obtained with a reduced basis provides the same statistical properties than the exact calculation. 

We should stress that, as the number of particles increases, the exact quantum ground state tends to the one obtained with the reduced basis ${\cal B}_\sigma^\kappa[0,0]$.

\section{Concluding remarks}\label{conclusion}

In this work, we built a sequence of ever approximating bases for the infinite-dimensional Hilbert space $\mathcal{H}$ of matter interacting with radiation. The ground state (which is the one under study) obtained in these truncated spaces differs from the exact ground state by no more than a certain allowed error ${\rm e}_{\rm rr}$ as measured by the fidelity $F$ between the two states. We have shown examples for both ${\rm e}_{\rm rr} = 10^{-10}$ and ${\rm e}_{\rm rr} = 10^{-15}$. The reduced bases provide solutions with the same statistical properties as those of the exact solution, and are especially useful when the number of particles is large. In fact, at different orders of the approximation one may study the physical properties of the system for any number of particles, as was exemplified even for a single particle in~\cite{cordero19}.

\appendix

\section{Dimensions of the \textsc{RWA} Bases .}\label{ap.dim}

For the $3$-level atoms interacting dipolarly with two modes of electromagnetic field, the degeneracy of states with fixed values $\kappa=\{k_1,k_2\}$ of the operators $\op{K}_1,\,\op{K}_2$ respectively (see Table~\ref{t.K3l}), is given by the dimension of the subspace ${\cal B}_\textsc{rwa}^{(\kappa)}$  Eq.~(\ref{eq.baseRWA}). To find the dimension as a function of the parameters we proceed as follows: for a fixed set of values $\{N_a,\,k_1,\,k_2\}$ the number of elements of the basis ${\cal B}_\textsc{rwa}^{(\kappa)}$ is calculated by means of equations of the form
\begin{equation}
a_1N_a^2 + a_2 N_a + b_1 k_1^2 + b_2 k_1 + c_1 k_2^2 + c_2 k_2 + d \, ,
\label{eqA1}
\end{equation}
Notice that an equation of second order in the variables $N_a,\, k_1$ and $k_2$ is proposed, because the states are the direct product of five Fock states, similar to a problem of five harmonic oscillators, whose degeneracy is given by a second order equation.

By comparing expression~(\ref{eqA1}) with the dimension found from the basis, given (seven sets of) values for $N_a,\, k_1$ and $k_2$, the coefficients $a_j,\,b_j,\,c_j$ and $d$ are determined and provide us with an analytical expression for the dimension. The region of the validity of the expression is also obtained.

After this procedure is finished  the expression for the dimension of the basis for each atomic configuration and any value of $\{N_a,\,k_1,\,k_2\}$ is obtained.

For the $\Lambda$-configuration one gets
\begin{widetext}
\begin{equation}\label{eq.dimL}
{\rm dim}\left({\cal B}_{\textsc{rwa}}^{(\kappa)}\right)= \left\{ \begin{array}{l l} 
g(N_a+k_1-k_2+1)& N_a<k_2\ \wedge\  k_1\leq k_2\\[3mm]
g(k_1+1)&k_2\leq N_a\ \wedge\  k_1\leq k_2 \\[3mm]
g(k_2+1)& k_2\leq N_a\ \wedge \ k_2<k_1 \\[3mm]
g(N_a+1)&\textrm{other case}  \end{array}\right.\,,
\end{equation}
where we defined
\begin{equation}
g(x):= \frac{(x)(x+1)}{2}\,,
\end{equation}
and the operators $\op{K}_1,\,\op{K}_2$ are given in Table~\ref{t.K3l}. Given the values $k_1=0,\,1,\,2,\,\dots$, the values of $k_2$ are limited by $k_2=0,\,1\,\dots,\,k_1+N_a$; in all other cases the subspace is empty.

For the $\Xi$-configuration one finds
\begin{equation}\label{eq.dimXi}
{\rm dim}\left({\cal B}_{\textsc{rwa}}^{(\kappa)}\right)=\left\{\begin{array}{l l}
g(k_1-k_2+1)& k_1-k_2\leq N_a\ \wedge\ k_1\leq 2\,k_2\\[3mm]
g(k_1+1)-g(k_1-k_2) - 2 \, g(k_2)& k_1-k_2\leq N_a\ \wedge\ 2\,k_2 < k_1\\[3mm]
g(N_a+1) - g(N_a-k_2)& N_a<k_1-k_2\ \wedge\ k_2<N_a \\[3mm]
g(N_a+1)& \textrm{other case}\,.
\end{array}
\right.\,,
\end{equation}
In similar way to the $\Lambda$-configuration, the values satisfy $k_1=0,\,1,\,2,\,\dots$, and $k_2=0,\,1,\,\dots,\,k_1$.

For the $V$-configuration
\begin{equation}\label{eq.dimV}
{\rm dim}\left({\cal B}_{\textsc{rwa}}^{(\kappa)}\right)=\left\{\begin{array}{l l}
(k_2+1)(k_1-k_2+1)& k_1\leq N_a\\[3mm]
&(2N_a \leq k_1 \ \wedge\ k_2\leq N_a)\, \\[-1mm] h(k_2,N_a)& \\[-1mm] &\vee\, (N_a<k_1 \ \wedge\ k_1 < 2N_a\, \wedge\ N_a < k_1-k_2)\\[3mm]
&(N_a<k_1\ \wedge\  k_1<2N_a\ \wedge\ N_a<k_2)\, \\[-1mm] h(k_1-k_2,N_a) & \\[-1mm] &  \vee\, (2N_a\leq k_1\ \wedge\ k_1-k_2\leq N_a)\\[3mm]
(k_1-k_2)k_2 + h(N_a,k_1)-g(k_1)&N_a<k_1\ \wedge\ k_1< 2N_a\ \wedge\ k_2\leq N_a\ \wedge\  k_1-k_2\leq N_a\\[3mm]
g(N_a+1)&\textrm{other case}
\end{array}\right.\,,
\end{equation}
\end{widetext}
with $k_1=0,\,1,\,2,\,\dots$, and $k_2=0,\,1,\,\dots,\,k_1$ and where we defined
\begin{equation}
h(x,y):= (x+1)(y+1)-g(x)\,.
\end{equation}
to simplify the notation.
\vspace{0.1in}

\section*{Acknowledgments}
This work was partially supported by DGAPA-UNAM under projects IN101619 and IN101217.

\end{document}